\documentclass{PoS}

\title{Narrow-line Seyfert 1 galaxies in the context of the Quasar Main Sequence}

\ShortTitle{NLS1 and the Quasar Main Sequence}

\author{\speaker{Bo\. zena Czerny}\\
        Center for Theoretical Physics, Polish Academy of Sciences, Al. Lotnik\' ow 32/46, Warsaw, Poland;
        E-mail: \email{bcz@cft.edu.pl}}
\author{Swayamtrupta Panda\\
        Center for Theoretical Physics, Polish Academy of Sciences, Al. Lotnik\' ow 32/46, Warsaw, Poland}
\author{Marzena \' Sniegowska\\
        Center for Theoretical Physics, Polish Academy of Sciences, Al. Lotnik\' ow 32/46, Warsaw, Poland}
\author{Szymon Koz\l owski\\
        Warsaw University Observatory, Al. Ujazdowskie 4, Warsaw, Poland}
\author{Marek Niko\l ajuk\\
        Faculty of Physics, University of Bia\l ystok, ul. Ciolkowskiego 1L, Bia\l ystok, Poland}
\author{Pu Du\\
        Institute of High Energy Physics, Chinese Academy of Sciences, Beijing, China}
\author{Bei You\\
  School of Physics and Technology, Wuhan University, Wuhan, China}

\abstract{Narrow-line Seyfert 1 galaxies are defined on the basis of their line widths, and they are generally considered to be high Eddington ratio sources. But in the context of the Quasar Main Sequence, high Eddington rate sources are those which have weak [O III] lines and strong Fe II lines. There is an overlap between the two populations, but they are not identical. Thus these two selection criteria give a different view on which objects are actually high Eddington ratio sources. We discuss this issue in the context of the broad band spectral energy density, emission line shape modeling, Fe II pseudo-continuum strength, and the level of X-ray variability. We also discuss the issue of the viewing angle and the insight one can gain from spectropolarimetric observations. We conclude that there is no single driver behind the Quasar Main Sequence, and the Eddington ratio of the source cannot be determined from the source location on the optical plane alone. On the other hand, an expected range of AGN parameters combined with a simple model of the Fe II production represent well the observed coverage pattern of the plane, with not much effect needed from the dispersion due to the viewing angle.}

\FullConference{Revisiting narrow-line Seyfert 1 galaxies and their place in the Universe - NLS1 Padova\\
		9-13 April 2018 \\
		Padova Botanical Garden, Italy}

\begin{document}

\section{Location of the high Eddington sources in the QMS optical plane}

The concepts of the narrow-line Seyfert 1 (NLS1) and of Quasar Main Sequence (QMS) were introduced several years ago in an attempt to organize active galaxies despite variety of their properties, and with the aim to look for the most important physical parameters.  

NLS1 class was defined by [30] as active galaxies with the Full Width Half Maximum (FWHM) of H$\beta$ line narrower than 1000 km s$^{-1}$, and with the equivalent width (EW) ratios of [OIII]5007\AA~to H$\beta$ smaller than 3. Later this limit has been modified, most frequently to 2000 km s$^{-1}$ (e.g. [31], [37]). It was argued [30] that these sources differ strongly from the typical type 2 Seyfert galaxies by having dense Broad Line Region (BLR) in addition to Narrow Line Region (NLR), and thus represent the tail of the trend of decreasing EW(H$\beta$) with increasing FWHM(H$\beta$). NLS1 were interpreted as high Eddington ratio sources by [33] on the basis of a huge soft X-ray excess observed in the NLS1 galaxy RE J1034+396.

The concept of the QMS was introduced with [7] where the authors used the Principal Component Analysis (PCA) to simplify the study of the correlations between several measured properties in a sample of 87 objects. Finally, they selected 13 measured quantities as basic, including the ratio $R_{Fe}$ of the EW of optical Fe II line to EW(H$\beta$). All measured quantities are well aligned along the Eigenvector 1 (EV1), and this correlation was later generalized to include the X-ray properties [8]. Later many studies concentrated just on the optical plane, i.e. on the relation $R_{Fe}$ vs. FWHM(H$\beta$) ([36]; see [25] and the references therein), one of the relations already studied by [7]. [7] also proposed an interpretation of the result suggesting that the Eddington ratio is the physical driver behind their Eigenvector 1. Within this concept, high Eddington ratio sources are located at the end of the sequence, being strong Fe II emitters in the optical band.

In [7], their Table 3, the authors listed the correlation of the EV1 with the contributing quantities, and it was seen that the QMS was predominantly built on two quantities: $R_{Fe}$ and FWHM(H$\beta$) as in their sample both quantities strongly correlated with EV1 (negatively and positively; respectively). Thus in subsequent papers the complex PCA analysis was frequently replaced with the study of the optical plane of EV1: plots of FWHM(H$\beta$) against $R_{Fe}$ ([36], [24], [34], [25]). This reduction is not likely to affect the main conclusion of [7] that high Eddington ratio sources should be among the strong Fe II emitters. 

\begin{figure}
\includegraphics[scale=0.75]{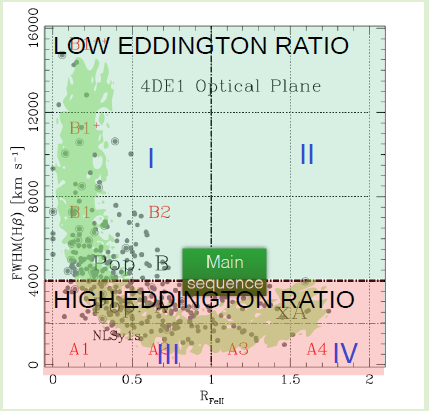}
\includegraphics[scale=0.75]{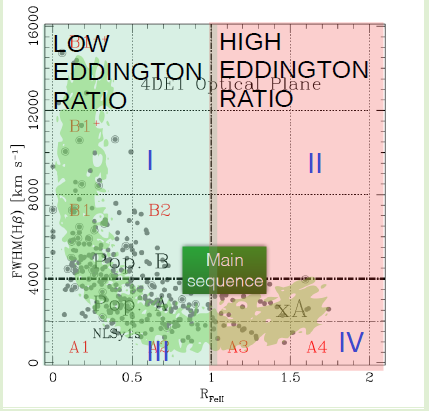}
\caption{The schematic location of the high Eddington ratio sources on the optical plane (pink shadow) and the low Eddington ratio sources (green shadow) according to the interpretation based on line width (left panel) and Fe II intensity (right panel); the underlying figure comes from [25]. 
\label{fig:problem}}
\end{figure}

This, however, leads us to a disagreement between the concept that AGN with very narrow H$\beta$ lines are all high Eddington sources, independent from their emissivity. The problem is illustrated in Fig.~\ref{fig:problem}. If the line width itself determines the Eddington ratio then high Eddington ratio sources should opulate regions III and IV. If, however, high Eddington ratio is predominantly related to the Fe II strength then high Eddington ratio sources should populate regions II and IV. 

\section{The dependence of the QMS in the optical plane on the black hole mass}
\label{sect:BHmass}

QMS is an attractive concept, introducing an order in a quasar variety. Even if the study concentrates on type 1 AGN, where the view towards the nucleus is unobscured, there are systematic differences in the shapes of emission lines and the line ratios. This was the reason why the NLS1 class of AGN received considerable attention. However, the studies of quasars showed that actually the qualitative change in the line shapes happens at 4000 km s$^{-1}$ ([36]) instead of 2000 km s$^{-1}$. This is most likely connected with the systematic difference between the black hole masses of the sources: NLS1 masses are systematically smaller than masses in distant quasars. 
If the division between the classes is indeed based on the fixed Eddington ratio, then the transition between the classes expressed as a value of FWHM should indeed depend on the black hole mass. The measured line width, FWHM, is related to the black hole mass and the distance to the BLR through a virial theorem
\begin{equation}
FWHM = \sqrt{GM \over R_{BLR}},
\end{equation}
where we neglected the issue of the virial factor. We can reasonably assume that the physical transition between the two types of objects takes place at a specific value, $f_{tr}$, of the Eddington ratio, $f_{tr}=L_{bol}/L_{Edd}$. Now the exact quantitative prediction depends on the assumption whether the distance to the BLR scales with the ionizing flux and bolometric luminosity $R_{BLR} \propto L^{1/2}_{bol}$ (e.g. [38]) or with the monochromatic flux $R_{BLR} \propto L_{\nu}^{1/2} \propto (M \dot M)^{2/3}$ (see e.g. [5] for data points and [10] for arguments why this is quite likely). We thus get the prediction for the transition value $FWHM_{tr}$ between low and high Eddington ratio sources  


\begin{eqnarray}
FWHM_{tr} & \propto M^{1/4} & ~~~~~{\rm if} ~~ R_{BLR} \propto L_{bol}^{1/2}, \\
FWHM_{tr} & \propto M^{1/6} & ~~~~~{\rm if} ~~ R_{BLR} \propto L_{\nu}^{1/2}.
\end{eqnarray}

The change of the mean black hole mass from $10^7 M_{\odot}$ to $3 \times 10^8 M_{\odot}$ would thus imply the change of the typical FWHM limit dividing the two populations by a factor of  2.3 in the first case, and 1.8 in the second case. This discussion illustrates that the QMS indeed cannot be expected to be very narrow since a number of basic parameters of an active nucleus can affect the properties: the black hole mass, accretion rate, spin, and the viewing angle. 
However, if the argument above is correct, it would be consistent with the vertical division between the low and high Eddington ratio on the optical plane. On the other hand, the non-uniform coverage of the optical plane may introduce an impression of the vertical division, if the upper right corner (region II; $FWHM > 4000$ km s$^{-1}$, $R_{Fe} > 1$) and left lower corner (region III; $FWHM < 4000$ km s$^{-1}$, $R_{Fe} < 1$)of Fig.~\ref{fig:problem} are indeed empty. However, region III in Fig. 1 is not empty, and region II is also not empty, if all objects are included, as discussed below.

\section{Location of the AGN with specified Eddington ratio in the optical QMS plane}

\begin{figure}
\includegraphics[scale=0.4]{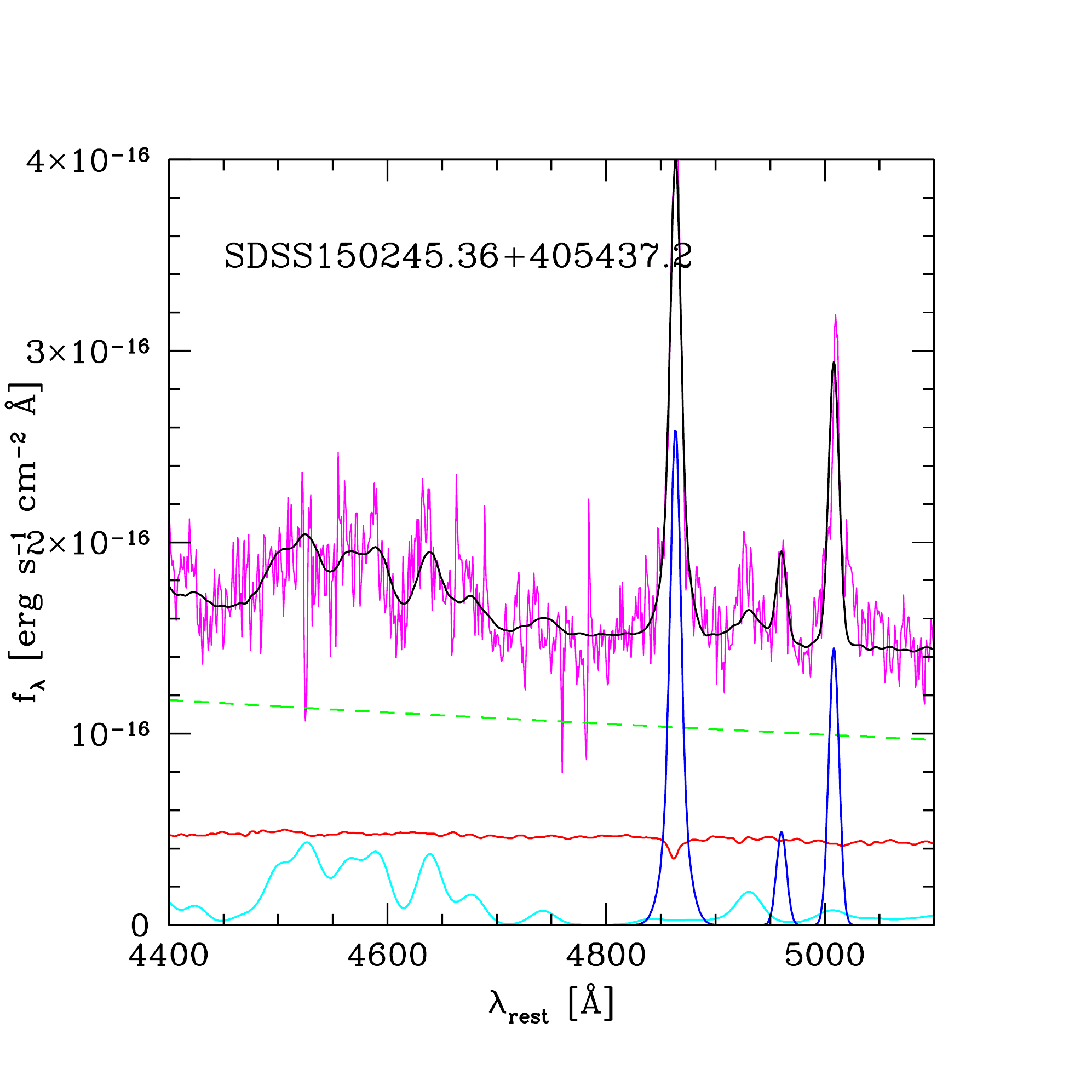}
\includegraphics[scale=0.4]{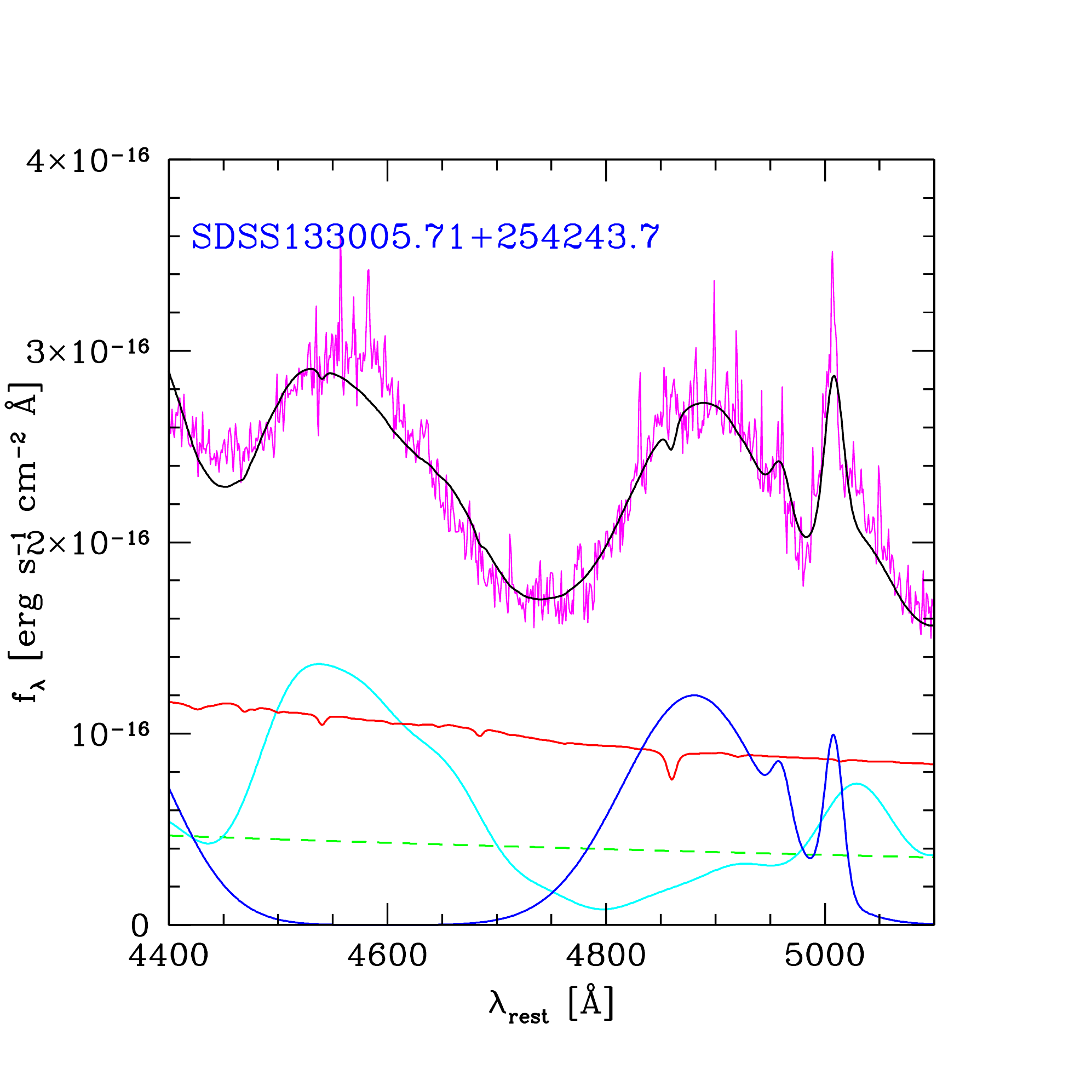}
\caption{Two examples of the extremely strong Fe II emitters from \' Sniegowska et al. (2018): left panel shows a source with very narrow lines and right panel shows an example with very broad lines. Magenta lines show the data, black line the model fits, and other lines show the fit components (power law - green dashed line, starlight - red line, Fe II pseudo-continuum - cyan, emission lines - blue). 
\label{fig:Marzena}}
\end{figure}

The upper right corner and left lower corner of Fig.~\ref{fig:problem}, however, are not empty. 
This is certainly not true for the left lower corner: already [30] stressed that of their extremely narrow NLS1 objects some are strong Fe II emitters while some (like Mrk 359, 783 and 1126) are quite weak.  This also is not true for the upper right corner: in our study (\' Sniegowska et al. 2018) of 27 extreme Fe II emitters with high quality data from SDSS quasar catalog [35] we found that of the six most extreme emitters three have very narrow lines (FWHM below  2100 km s$^{-1}$) and high Eddington ratio but the remaining three have broad lines (above 5400 km s$^{-1}$) and low Eddington ratios. Two examples are shown in Fig.~\ref{fig:Marzena}.

\begin{figure}
\includegraphics[scale=0.4]{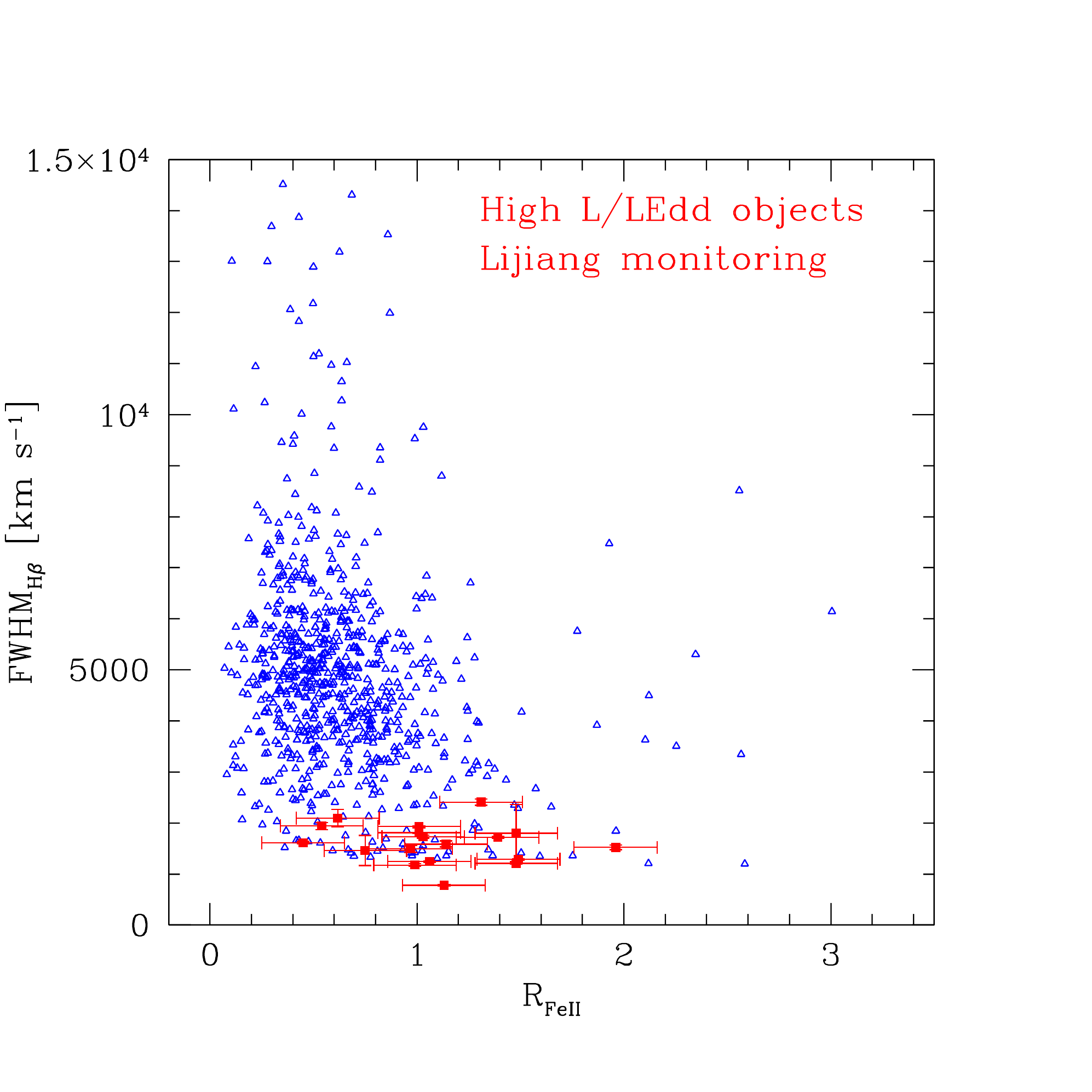}
  \caption{The location of the super-Eddington sources selected for reverberation measurements [15] on the  optical plane (red points with errorbars), blue points are quasars from \' Sniegowska et al. (2018). 
\label{fig:Pu}}
\end{figure}

Super-Eddington sources ([16], and the references therein) also do not seem to be located at the right hand side of the plot: instead, they cover a broad range of $R_{Fe}$ values between $\sim 0.5$ and $\sim 2.0$, with the average values of $\sim 1$ (see Fig.~\ref{fig:Pu}). On the other hand, the measurement of the Eddington ratio in a given object is not simple. The problem is associated with the measurement of both the bolometric luminosity and of the black hole mass.

The measurement of the bolometric luminosity requires an integration over the broad band spectrum. Such a spectrum is not always available, and in addition most of the AGN luminosity is emitted in the unobserved far-UV band. Frequently used constant bolometric correction to the measured optical flux (a factor of 9\footnote{https://ned.ipac.caltech.edu/level5/March04/Risaliti/Risaliti2\_5.html}) is a very crude approximation.  

\subsection{Direct measurements of the black hole mass in NLS1}

It is thus crucial, particularly in the case of NLS1, to find reliable mass measurement methods. However, this is not simple. Standard methods, based on the FWHM, require the knowledge of the virial factor (e.g. [32], [9], [5]) which itself seems to be coupled to the line width [26]. If several black hole mass measurements are used for a single source, the result can be quite inconclusive. For example, the black hole mass measurement in the NLS1 galaxy RE J1034+396 with methods based on H$\beta$, Mg II and [OIII] line width, X-ray excess variance, stellar dispersion, broad band spectrum fitting, and frequencies of the Quasi-periodic oscillations firmly detected in X-ray emission from this source give the mass range from well below $10^6 M_{\odot}$ to above $10^7 M_{\odot}$ [11]. Reverberation mapping measurements for this source currently performed (Jian-Min Wang, private communication) may solve the problem of this particular source, but in general, the problem is far from solved. For example, otherwise powerful method of the X-ray excess variance gives the values of the black hole mass reliably (e.g.  [29], [20]), but as shown in [28] some of the NLS1 (but not all!) are clearly outliers from the general scaling of the variability with mass.

\subsection{NLS1 impostors}

There is a small group of sources with location on the optical plane strongly influenced by the extremely low viewing angle. Since the BLR is certainly rather flat, lines observed along the symmetry axis would display very narrow shapes, and thus would be located in the much lower part of the optical plane than expected. The best way to determine the true Keplerian velocity in such objects is through the spectropolarimetric observations ([1], [2]; see also Popovic, this volume). In the polarized light lines (measurements are made for H$\alpha$) in such objects are much broader than seen in unpolarized light since most scattering happens close to the equatorial plane and the full range of Keplerian velocities is revealed.  

\subsection{The X-ray spectral slope and the Eddington ratio}

Important information can be gained from the X-ray measurements since we know from the numerous studies of the Galactic black holes that sources at high Eddington ratios usually show strong soft component either in the form of steep X-ray spectra or clearly disk-dominated spectra with a weak hard X-ray tail [14]. Weak Fe II emitters always have hard spectra, but strong Fe II emitters can be either steep or flat in the ROSAT data, but they are always X-ray weak [22]. If this is generally true it might rather point out toward vertical division in Fig.~\ref{fig:problem} but it is not consistent with the sources discussed in the previous paragraph. Unfortunately, for many objects, including the sources discussed in Sec.~\ref{sect:BHmass} we do not have X-ray spectra.

\section{Theoretical model of the QMS optical plane}
\label{sect:theory}

\begin{figure}
\includegraphics[scale=0.6]{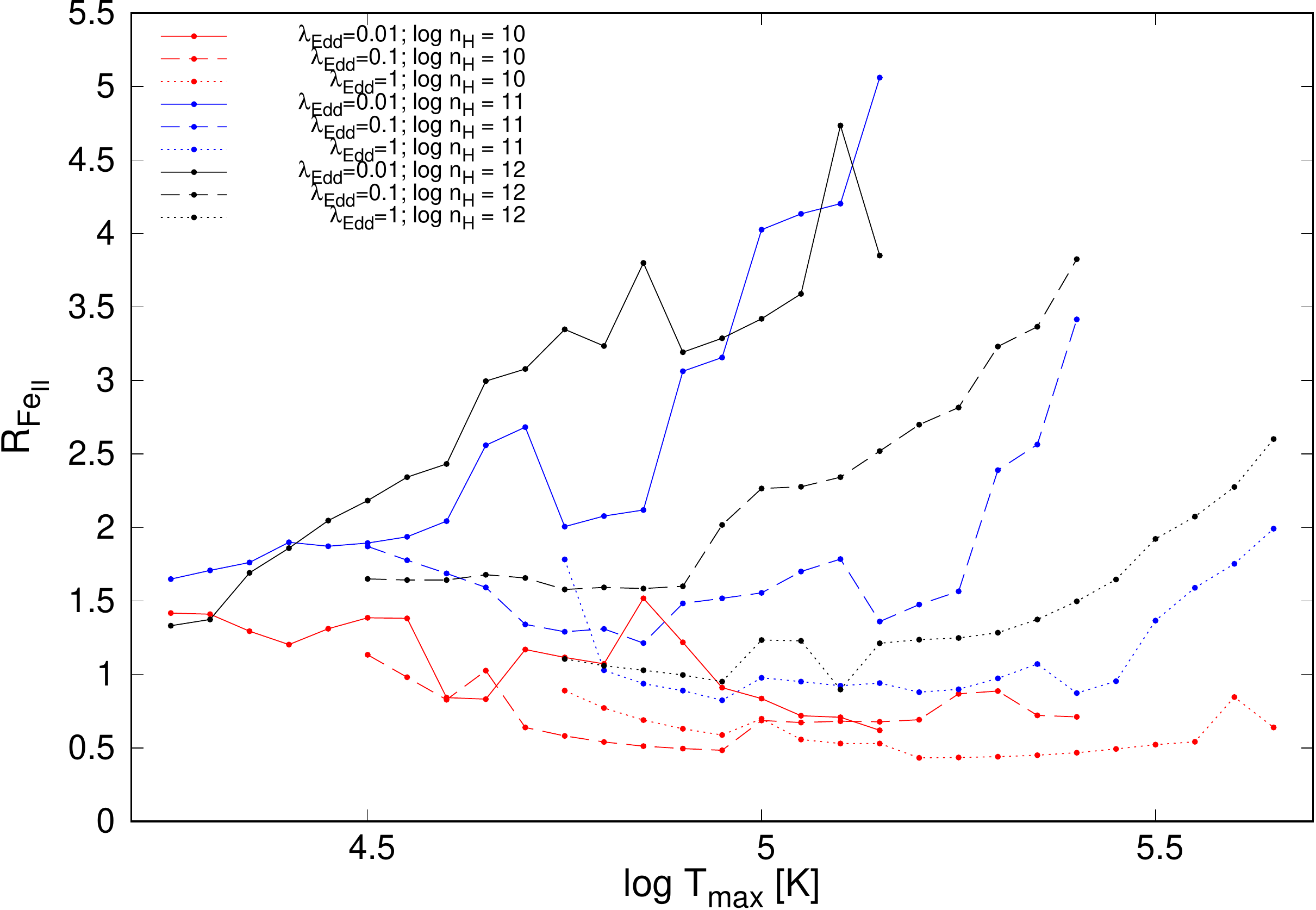}
\caption{The preliminary results for the dependence of the relative efficiency of the Fe II to H$\beta$ production modeled with CLOUDY  as a function of cloud density, Eddington ratio and the maximum temperature of the accretion disk (Panda et al., in preparation). 
\label{fig:Swayam1}}
\end{figure}

Since the direct observational approach to the QMS optical plane appears complex, we address the issue from the theoretical point of view. We perform simulations of the Fe II and H$\beta$ production in the BLR. We assume a single-zone approximation, constant density clouds, and no shielding effect of the clouds themselves. The total surface density of the clouds is fixed at $N_H = 10^{24}$ cm$^{-2}$. The incident spectrum is modeled as consisting of a standard Shakura-Sunyaev Keplerian disk (Shakura \& Sunyaev 1973) with an additional hard X-ray power law contribution. The relative normalization between the UV disk component and the X-ray component is taken from [23], the X-ray slope is fixed at the photon index 1.9. The distance of the clouds from the black hole is calculated from the formula for the BLR radius of [5]. Thus, our model has three arbitrary parameters, and we express them as: the maximum temperature of the accretion disk, $T_{max}$, the Eddington ratio, and the cloud local density, $n$. 

\begin{figure}
\includegraphics[scale=0.6]{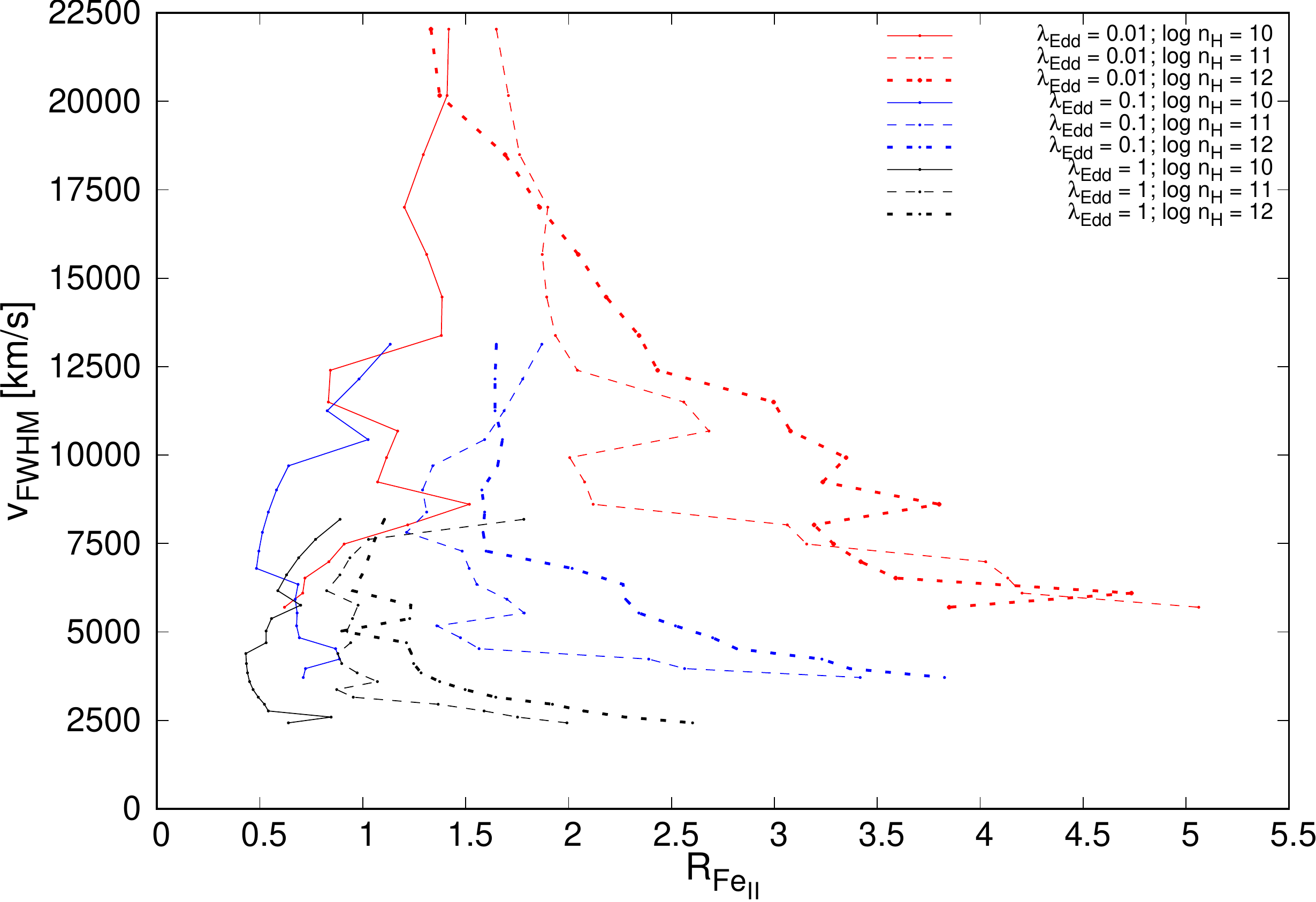}
\caption{The preliminary results for the coverage of the optical plane by the model used in Fig.~\ref{fig:Swayam1} (Panda et al., in preparation). 
\label{fig:Swayam2}}
\end{figure}

~
Computations of the line emission from the model are performed with the CLOUDY code, version 13 [17]. The results of the luminosity ratios $R_{Fe}$ of Fe II to H$\beta$ are shown in Fig.~\ref{fig:Swayam1}. We see that the model represents reasonably well the average values in the high quality sub-sample of the Shen et al. (2011). If only sources with the measurement errors of Fe II and H$\beta$ below 20\% are selected, the average Eddington ratio is 0.1, the average black hole mass is $\log M = 8.4$ in solar units, which corresponds to $\log T_{max} = 4.80$. The mean and median values of $R_{Fe}$ in this subsample are 0.64 (\' Sniegowska et al. 2018), and in the model we obtain values from 0.7 to 1.1, depending on the cloud density. Thus, on average, the Fe II production in the clouds does not require additional turbulent heating, radiative  processes are efficient enough, and radiatively driven Fe II production is consistent with reverberation mapping of Fe II [19]. The time delay measured to the Fe II production region is not very different from time delay measured for H$\beta$ line. And more
interesting, the time lag ratio of Fe II and H$\beta$ has a strong correlation with $R_{Fe}$
([3], [19]). This correlation may be worth consideration in the future.
From our current single-zone modeling we see that $R_{Fe}$ is not a monotonic function, either of $T_{max}$, or of $L/L_{Edd}$. High Fe II emitters in the model can be found also among the low Eddington ratio sources. However, the efficient Fe II production requires high cloud densities, as already discussed by [6].

Since our model also determines the line width, we thus located the theoretical points directly on the optical plane (see Fig.~\ref{fig:Swayam2}). 
The plane is well covered apart from the strong Fe II emitters tail. High density material ($10^{12}$ cm$^{-3}$) is needed to represent a significant part of the plane. The need for high densities in the outflowing wind has already been concluded in the context of EV1 by [22], although actual dynamics of the region is not well known, and the possibility of an inflow (e.g. [18], [39]), a failed wind ([10], [11], [13]), or a static atmosphere of an accretion disk are under consideration [4]. High densities were also favored by Adhikari et al. (2018) to explain the smooth profiles in NLS1, without a clear division into BLR and NLR due to the presence of the dust in the intermediate region seen in lower density modeling [27]. 

The model shows some overlap of high and low Eddington ratio points in the strong Fe II emitters' region which is consistent with the findings of \' Sniegowska et al. (2018). However, we have to keep in mind that the current simulations were based on the Shakura-Sunyaev disk model which may not be valid, particularly in the case of NLS1. We plan to expand the study presented here to more general shapes of the incident continuum, including the warm corona (Panda et al., in preparation).    

\section{Discussion}

Quasar optical plane does not seem to be driven by a single physical parameter, like the Eddington ratio or the maximum disk temperature (or, equivalently, the peak of the spectral energy distribution (SED). Therefore, it is indeed impossible to identify uniquely a location of high Eddington ratio sources on the QMS optical plane and a single parameter. This is consistent with the simple fact that even a stationary accreting black hole has a number of independent parameters: black hole mass, spin, accretion rate and a viewing angle, and the dispersion in these parameters in AGN leads to a complex coverage of the plane. The predictions based on the model of Fe II production are roughly consistent with the plane coverage by the observational points. It is important to note that the model discussed in Sect.~\ref{sect:theory} does not include the viewing angle although it is in principle important for the measurement of the continuum flux (the flat disk luminosity should be proportional to cosine of the viewing angle). However, if the H$\beta$ and Fe II lines originate roughly from the same region, the effect of the viewing angle should disappear. Instead, it should reappear in the measurement of the line width since the BLR is expected to be also flattened. Our optical plane coverage in the vertical direction is already well covered so apparently the role of the viewing angle is not essential and most of the dispersion in FWHM(H$\beta$) apparently comes from the dispersion in the black hole mass and accretion rate, as already stressed in [9]. Observations also indicate that even highly inclined sources have rather typical line shapes, as was shown by [40] on the basis of a sample of lobe-dominated quasars.   

Those results imply that a selection of sources close to Eddington luminosity cannot be done at the basis of just the QMS optical plane, as discussed by [25]. Other parameters, like soft-X-ray slopes may provide additional information since the optical plane contains only a part of the information about the source. Other wavelengths contain additional information and may show a different pattern.

For example, instead of using the optical plane, we can also study QMS in the UV plane, using Mg II line instead of H$\beta$ and Fe II$_{UV}$ instead of Fe II$_{opt}$. The plane coverage looks apparently similar if plotted in the linear scale but if plotted in log-log scale it shows much stronger coupling between the measured quantities. Further inspection shows that the leading correlation is actually between FWHM(Mg II) and EW(Mg II) while Fe II$_{UV}$ is uncorrelated (\' Sniegowska et al, in preparation). This is consistent with the weak coupling between Fe II emission in the optical and in the UV band [21].

\section*{Acknowledgements}

This conference has been organized with the support of the
Department of Physics and Astronomy ``Galileo Galilei'', the 
University of Padova, the National Institute of Astrophysics 
INAF, the Padova Planetarium, and the RadioNet consortium. 
RadioNet has received funding from the European Union's
Horizon 2020 research and innovation programme under 
grant agreement No~730562. BC, SP, and MS acknowledge the financial support by National Science Center grant Nr. 2015/17/B/ST9/03436/. SK acknowledges the financial support of the Polish National Science Center through the OPUS grant 2014/15/B/ST9/00093. The project was also partially supported by by the Polish Ministry of Science and Higher Education under
subsidy for maintaining the research potential of the Faculty of Physics, University of Bia\l ystok.


\begin{thebibliography}{99}
\bibitem[1]{} Afanasiev, V.L., Popovic, L.C., 2015, ApJ, 800, L35
\bibitem[2]{} Baldi, R.D., Capetti, A., Robinson, A., Laor, A., Behar, E., 2016, MNRAS, 458, L69
\bibitem[3]{} Barth, A.J. et al., 2013, ApJ, 769, 128
\bibitem[4]{} Baskin, A., Laor, A., 2018, MNRAS, 474, 1970
\bibitem[5]{} Bentz, M.C. et al., 2013, ApJ, 767, 149
\bibitem[6]{} Bruhweiler, F., Verner, E., 2008, ApJ, 675, 83
\bibitem[7]{} Boroson, T.A., Green, R.F., 1992, ApJS, 80, 109
\bibitem[8]{} Brandt, W.N., Boller, Th., 1998, AN., 319, 7
\bibitem[9]{} Collin, S., Kawaguchi, T., Peterson, B.M., Vestergard, M., 2006, A\&A, 456, 75
\bibitem[10]{} Czerny, B., Hryniewicz, K., 2011, A\&A, 525, L8
\bibitem[11]{} Czerny, B. et al., 2016a, A\&A, 594, A102
\bibitem[12]{} Czerny, B. et al., 2015, AdSpR, 55, 1806
\bibitem[13]{} Czerny, B. et al., 2016b, ApJ, 832, 15
\bibitem[14]{} Done, C., Gierlinski, M., Kubota, A., 2007, A\&ARv, 15, 1
\bibitem[15]{} Du, Pu et al., 2016, ApJ, 825, 126
\bibitem[16]{} Du, Pu et al., 2018, ApJ, 856, 6
\bibitem[17]{} Ferland, G.J. et al., 2013, RMxAA, 49, 137
\bibitem[18]{} Ferland, G.J., et al., 2009, ApJ, 707, L82
\bibitem[19]{} Hu, C. et al., 2015, 804, 138
\bibitem[20]{} Kelly, B.C. et al., 2013, ApJ, 779, 187
\bibitem[21]{} Kovacevic-Dojnovic, J., Popovic, L., 2015, ApJS, 221, 35
\bibitem[22]{} Lawrence, A., Elvis, M., Wilkes, B.J., McHardy, I., Brandt, N., 1997, MNRAS, 285, 879
\bibitem[23]{} Lusso, B., Risaliti, G., 2017, A\&A, 602, A79
\bibitem[24]{} Marziani, P., Sulentic, J.W., Zwitter, T., Dultzin-Hacyan, D., Calvani, M., 2001, ApJ, 558, 553
\bibitem[25]{} Marziani, P. et al., 2018, FraSS, 5, 6
\bibitem[26]{} Mejia-Restrepo, J.E.,  Lira, P., Metzer, H., Trakhtenbrot, B., Capellupo, D.M., 2018, NatAs, 2, 63
\bibitem[27]{} Netzer, H., Laor, A., 1993, ApJ, 404, L51
\bibitem[28]{} Nikolajuk, M., Czerny, B., Gurynowicz, P., 2009, MNRAS, 394, 2141
\bibitem[29]{} Nikolajuk, M., Papadakis, I.E., Czerny, B., 2004, MNRAS, 350, L26
\bibitem[30]{} Osterbrock, D.E., Pogge, R.W., 1985, ApJ, 297, 166
\bibitem[31]{} Osterbrock, D.E., Pogge, R.W., 1987, ApJ, 323, 108 
\bibitem[32]{} Peterson, B.M. et al., 2004, ApJ, 613, 282
\bibitem[33]{} Pounds, K.A., Done, C., Osborne, J.P., 1995, MNRAS, 277, L5
\bibitem[34]{} Shen, Y., Ho, L.C., Natur, 513, 210
\bibitem[35]{} Shen, Y. et al., 2011, ??? I DALEJ NUMERACJA
\bibitem[35]{} \' Sniegowska, M. et al., 2018, A\&A, 613, A38
\bibitem[36]{} Sulentic, J.W., Zwitter, T., Marziani, P., Dultzin-Hacyan, D., 2000, ApJ, 536, L5
\bibitem[37]{} Veron-Cetty, M.-P., Veron, P., Goncales, A.C., 2001, A\&A, 372, 730
\bibitem[38]{} Wandel, A., Peterson, B.M., Malkan, M.A., 1999, ApJ, 526, 579
\bibitem[39]{} Wang, J-M. et al., 2017, NatAs, 1, 775
\bibitem[40]{} Wildy, C., Czerny, B., Kuzmicz, A., 2018, ApJ (in press); arXiv:1805.06263
\end{thebibliography}
\end{document}